# SECURE DATA ACCESS IN CLOUD ENVIRONMENTS USING QUANTUM CRYPTOGRAPHY

S. Vasavi Venkata Lakshmi, Ziaul Haque Choudhury

Department of Information Technology & Computer Applications

Vignan's Foundation for Science, Technology & Research (Deemed to be University)

Guntur, AP, India.

vasavisangasani@gmail.com, ziaulms@gmail.com

**ABSTRACT**

Cloud computing has made storing and accessing data easier, but keeping it secure is a big challenge nowadays. Traditional methods of ensuring data may not be strong enough in the future when powerful quantum computers become available. To solve this problem, this research uses quantum cryptography to protect data in the cloud environment. Quantum Key Distribution (QKD) creates secure keys by sending information using quantum particles like photons. Specifically, we use the BB84 protocol, a simple and reliable way to make secure keys that cannot be stolen without detection. To protect the data, we use the Quantum One–Time pad (QOTP) for encryption and decryption, ensuring the data stays completely private. This study shows how these Quantum methods can be applied in cloud systems to provide a strong defense against hackers, even if they have access to quantum computers. The combination of QKD, BB84, and QOTP creates a safe and reliable way to keep data secure when it is stored or shared in the cloud. Using quantum cryptography, this paper provides a way to ensure data security now and, in the future, making cloud computing safer for everyone to store their data securely and safely.

## 1. INTRODUCTION

Cloud Computing has revolutionized the digital era, turning the internet into a boundless warehouse of computing resources. With just an internet connection, individuals and businesses can tap into vast computational resources that scale with their needs. This "pay-as-you-use" model has broken down technological barriers, empowering startups, enterprises, and even governments to innovate faster, collaborate globally, and manage operations efficiently. However, as the dependency on cloud platforms grows, so does the challenge of ensuring data security.

The problem with Existing research mainly highlights the Access control system that uses DNA and some algorithms in cloud computing is weak against security attacks and is not flexible, making it hard to manage and access data properly. It also struggles with scalability when handling large datasets [1]. In the existing study, their proposed methods are very slow because DNA similarity searches and encryption take a lot of time and computing power. Here, the access control systems also have issues as they cannot easily adapt when user roles or permissions change [2]. They proposed cloud-based data storage using DNA and AES algorithms [3], the AES encryption layers confuse the data, making it more difficult for an attacker to decrypt the message, but consumes more time. The existing proposed systems are weak and easily breakable the algorithms due to quantum cryptography.

To solve those problems, we have proposed Quantum Cryptography, which is designed as a secure framework using Quantum Key Distribution (QKD) for key generation, employing the BB84 protocol. For encryption and decryption purposes, we use the Quantum One-Time Pad (QOTP), which ensures that data is completely secure. Even if someone intercepts the encrypted data, they cannot decrypt it without the quantum key. This system makes data more secure and protects it from common cyberattacks, even those from upcoming, powerful quantum computers. As quantum computers evolve and become more powerful, our system stands ready to defend against even the most advanced threats. With this, we are paving the way for the new era of secure data access in the cloud, where organizations can store, process, and share their information with complete confidence.

The main Contribution of this proposal work:

- Proposed quantum cryptography QKD with BB84 protocol and QOTP algorithm to ensure data protection against classical and quantum attacks for cloud-based security.

The remainder of the paper is divided into many sections. The literature review is covered in Section 2, the proposal technique is covered Section 3, and the experimental result is covered in Section 4. Section 5 concluded with a presentation of future projects.

## 2. LITERATURE SURVEY

The literature survey highlights security challenges in cloud computing. The methods are quite complicated because they use many steps, such as DNA-EXOR, DNASF, and other rules, which may take extra time [1]. As more users and data owners join the system, the way the model improves data access times may also not work, leading to slower performance and higher resource use in larger cloud setups. The authors proposed EFSS_I and EFSS_II are two DNA similarity search models. with a focus on EFSS_I, leaks information about the similarity of queried messages, compromising user privacy and getting less accuracy and with a low performance [2]. Their study used an (i) access control-based model that was less secure due to the weak protection, and (ii) machine learning-based models used DPHE, CryptoDL Framework, POCC, and so on. (iii) watermarking-based models like ILD, LIME, GAHSW, and JWEC. Their study highlights that no individual technique can guarantee complete data security, requiring the integration of multiple methods [3].

They used two hybrid algorithms DNA and AES systems these two algorithms were powerful but AES may be attacked via MITM (Man-in-Middle) or brute force. so it needs extra security to be safe [4]. The DML-DIV scheme may suffer from high communication overhead due to frequency verifications and significant computational overhead [5]. Data leakage prevention (DLP) and information leakage prevention (ILP) solutions rely heavily on technology and cannot effectively address non-technological threats, such as human errors or social engineering. Also, the DLP solutions struggle with unstructured data with unsupported formats and insider threats [6]. The protocol that was used could not protect the user's identity well. An attacker can figure out who a legitimate user is, which means the user's privacy is at risk. This protocol can be tricked by someone unknown persons and they can be pretending to be either a legitimate mobile user or the cloud server. This makes the system unsafe because the attackers could fake their identity and cause harm [7].

The existing study has some problems with finite precision in generating keystreams There are no established criteria for security and performance assessments, and hyperchaotic maps are used [8]. More fine-grained access control, or FGAC, together with certain algorithms, may be used to efficiently Preserve user data security, update user groups in real time, and limit access for fraudulent individuals [9]. The main drawback of it was poor flexibility and low accuracy. It might be easier to break the code if the same rules are always used to hide the picture, making the rules change based on the secret key or the picture itself could make it harder to use [10]. Their study [11], based on MDNAC leverages DNA's complexity for strong security when they practically implemented DNA-based cryptography and faced some challenges due to the complexities of DNA manipulation and sequencing. In [13], a new way of solving problems using DNA and computers is much faster because it uses special computer chips (GPUs). Algorithms like P-DNA have the advantage of maintaining better population diversity. But while exchanging the information between GPU and CPU The GPU's hardware limitations need to be taken into account. It provides a thorough analysis of data security, privacy, confidentiality, and integrity. challenges in cloud storage systems [14],[15]. Other studies [16] used three phases the first and second phases SHA512, dynamic DNA encoding for secret key generation, and the third phase for encrypting the audio. In their paper, they have large key space, providing strong defense against brute force attacks. The first is a description of a double-layer encryption of data concealment according to the suggested DNA approach and the DNA sequence [17].

When sending the secret message from the other end target organism may be challenging and potentially damage the organism. It talks about security and privacy [18], it uses the primary issue with SSGK, RSA, and verified secret sharing keys was that they neededboth forward and backward security for managing group keys, which called for certain protocols. [19] some algorithms to secure the data DNA stands (A, G, T, C) for security and this secured message can break through the quantum. The algorithm used was Steganography which provides excellent standard quality assessment findings and a high throughput capacity. A common security algorithm that works well for low-throughput real-time voice communication is modern-based cryptography. Its absence of standardized quality evaluation, including subjective and objective evaluation, raises concerns [20]. They talk about the BB84 protocol and how the key was generated through Quantum Key Distribution clearly by taking Alice and Bob's example in it. In its quantum cryptography, the inability of Eve to completely synchronize with Alice validates the artificial neural network-based key's security [21].

## 3. PROPOSED METHOD

The proposed method describes Quantum Cryptography, which secures data access in the cloud environment. By applying Quantum One–Time Pad (QOTP) data encryption and decryption is done and utilize Quantum Key Distribution (QKD) for secure key generation based on the BB84 protocol for communication. This framework ensures secure, scalable, and adaptable data access that offers strong defense against both classical and quantum threats. The suggested architecture is displayed in Fig. 1 below.

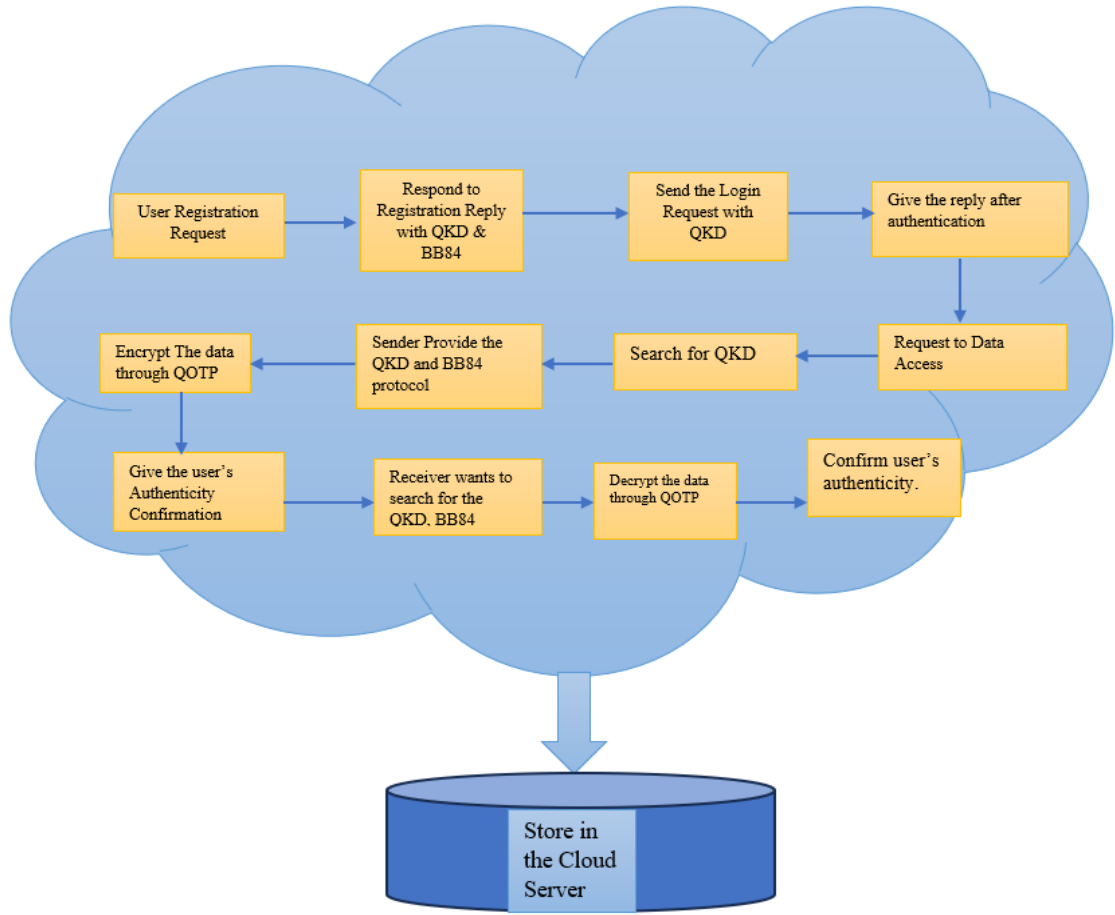

Fig. 1 The suggested design for cloud security using quantum cryptography

## 3.2 Quantum One-time pad

Frank Miller first gave a description of the Telegraphy one-time pad gadget security in 1882. Since the early 1900s, one-time pads have been utilized in certain situations. QOTP provides multiple security advantages over classical encryption techniques. One of its key benefits is unconditional security, as the randomness and uniqueness of the quantum key make it immune to brute-force attacks. Additionally, eavesdropping detection is a built-in feature of QKD, ensuring that any third-party interception of the quantum key is immediately noticed due to disturbances in quantum states. Furthermore, QOTP is quantum-resistant, meaning it remains secure even in the presence of quantum computers, unlike RSA or other public-key cryptographic methods that can be broken by Shor's algorithm. Given these properties, QOTP is considered an ideal solution for high-security applications such as cloud storage, military communication, financial transactions, and the development of a quantum-secure internet.

The sender uses the one-time pad encryption approach to encrypt the plaintext message by executing an XOR operation between the message and the randomly generated quantum key. Since the key is as long as the message and is completely random, the generated ciphertext is devoid of statistical patterns, guaranteeing total anonymity. A traditional communication channel is subsequently employed to transmit the encrypted information, and the receive with the same quantum key decrypts it with the same XOR operation to obtain the original plaintext. The process ensures that an attacker is unable to decipher the message in the absence of the precise quantum key even if h/she possesses unlimited processing capacity. Data encryption with quantum cryptography is demonstrated below in Figure 2.

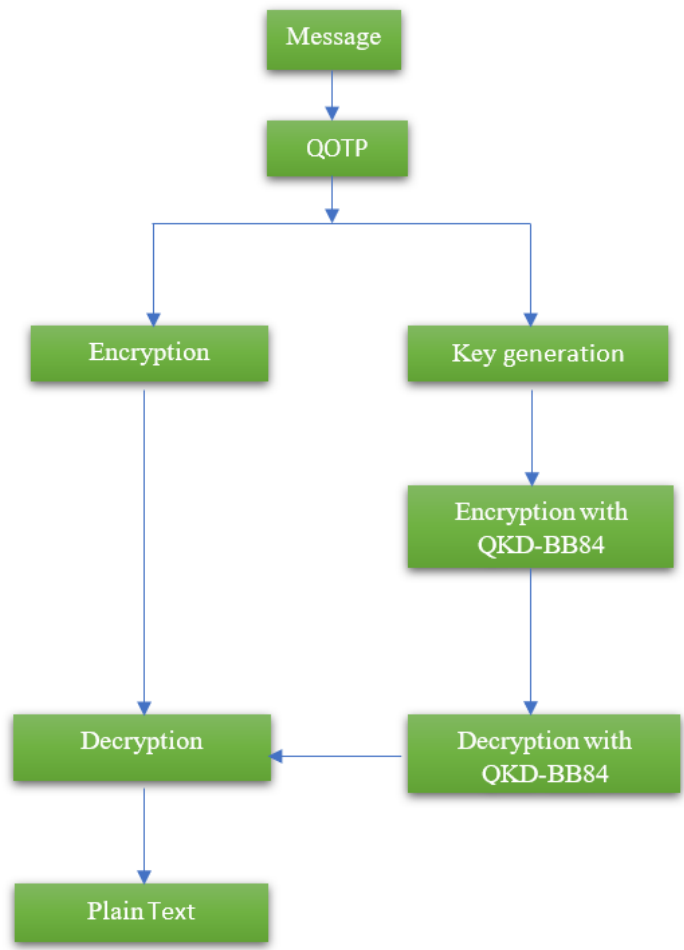

Fig. 2: The data encryption using quantum cryptography

In quantum cryptography, the quantum The unbreakable encryption technique known as one-time pad (QOTP) requires the use of a single-use pre-shared key that is at least as large as the message being sent. This method uses a random secret key, often known as a one-time pad, in conjunction with plaintext. Modular addition is then used to combine each bit or character of the plaintext with the matching bit or character from the pad in order to encrypt it [28].If all four of the following criteria are met, the resultant ciphertext cannot be decoded or broken:
1. 1. The plaintext's length and the key's length must match.
2. It is necessary to choose the key at random.

3. The key must never be partially or fully used again.
4. All communication parties must maintain complete confidentiality regarding the key.

In particular, one-time usage is essential. For instance, the following are the matching ciphertexts if two separate plaintext messages, p1 and p2, are both encrypted with the same key, k:

$$c1 = p1 \oplus k \qquad [1]$$

$$c2 = p2 \oplus k \qquad [2]$$

Here $\oplus$ means XOR, denotes the operation of bit by bit exclusive – or, such that if the attacker had the ciphertext c1 and c2, taking a simple XOR between c1 and c2 will result in the XOR between the two plaintexts $p1 \oplus p2$ .(The reason why is that applying XOR to k and itself gives an all – zeros bitstream.) $p1 \oplus p2$ is basically a running key cipher [29].

## 3.1 THE QUANTUM KEY DISTRIBUTION

An emerging key setup technique with unwavering security is called Quantum Key Distribution (QKD). To test an opponent, it uses Heisenberg's Uncertainty Principle or Bell's inequalities violation in entanglement-based methods. A quantum state is destroyed when it is measured according to Heisenberg's uncertainty-based methodology. generating the error detectable when conveying information on a quantum channel. Entanglement-based protocols make use of the fact that when an intruder attempts to measure the entangled quanta, information becomes real and leads to violations of Bell's inequalities. The theorem is also useful for detecting eavesdropping, as it prevents an attacker from copying quantum information [22] [23].

In a standard QKD procedure, a secret key is established by Bob, the receiver, and Alice, the transmitter, use both a classical and a quantum channel. The quantum channel can be an optical fiber or a free-space line-of-sight transmission, whereas the classical channel can be any conventional network link, such the internet or a phone network. Alice employs a laser source at the transmitting side to condition and broadcast individual polarized photons, commonly referred to as quantum bits or qubits. Bob observes qubits, photons, and other particles on the receiving end. or quantum bits. Bob watches as raw key bits are generated on the receiving side by single-photon detectors (SPDs). Common secret keys are generated once information is shared over the conventional route. Text, phone, and video messages containing sensitive data can be encrypted using these QKD secret keys [24].

In the QKD protocol, the useable key's length steadily gets shorter. Bob measures the polarized photons that Alice prepares as information. These techniques use the Quantum No-Cloning theorem and Heisenberg's Uncertainty principle to detect eavesdroppers based on transmission faults. A common prepare-and-measure-based QKD technique is BB84. A QKD systems normally consists of a quantum channel to send qubits, a classical channel to post-process, a single-photon source, a measurement of incoming qubits with detectors, and an authentication component to ensure the prevention of man-in-the-middle attacks. The combination of QKD with current cryptographic schemes adds security by offering a future-proof solution against quantum and classical attacks. QKD's eavesdropping detection capability, coupled with its dependence on quantum mechanics principles, makes it a secure means for protecting communications in cloud computing, financial transactions, and military uses.

## 3.3 BB84 PROTOCOL KEY SHIFTING

Brassard and Bennett created BB84., 1984 saw the introduction of the first quantum key distribution protocol [25]. However, it was implemented in 1991 to polarize single photons based on a random bit stream utilizing two bases. This process uses four states of polarization with two bases, one a rectilinear basis and the other a diagonal basis. According to the rectilinear basis, the logic 0 is a 0-polarized photon, the logic 1 is a 90-polarized photon, and the logic 1 is a 135-polarized photon. Table 1 shows the basis and the polarization states based on the bases.

Table 1 BB84 protocol key shifting

| Bit value | | 0 | 1 |
|---|---|---|---|
| Rectilinear Basis: | 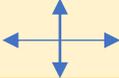 | 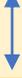 | 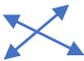 |
| Diagonal Basis: | 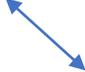 | | |

## 3.4 Quantum Transmission

Quantum transmission stage encompasses all the actions which are performed on quantum states, which includes communication across a quantum channel and the encoding and decoding of classical bits to quantum states.
1. Alice chooses a set of N classical bits X1,...XN that are evenly distributed.
2. Alice can choose the diagonal basis (D) or the rectilinear basis (R) from a series of random polarization bases. A measurement in one basis does not provide any information on a bit that is encoded in the other basis, demonstrating the unbiased nature of these two bases.
3. Alice successfully transfers her bit string into an assembly of photons that are polarized with respect to the chosen bases. When Bob gets the photons, he chooses at random (without Alice's help!) whether to measure each one in the diagonal or rectilinear basis to get classical bits. Both Alice and Bob then possess a classical bit string, which is represented by the notation X = (X1,…,XN) for Alice and Y = (Y1,...,YN) for Bob. We refer to this as the raw key pair.

## 3.5 SYSTEM COMPONENTS

- Cloud Environment: A cloud infrastructure where sensitive data is stored and must be accessed securely. The cloud acts as the server-side platform, hosting the data.
- User: An individual who wants to access data stored in the cloud. This user must authenticate themselves before accessing any sensitive information.
- Quantum Key Distribution (QKD): A technique that allows the used and cloud system to safely exchange encryption keys. This allows both parties to have the same secret key without an attacker being able to intercept it.
- BB84 Protocol: During the QKD process the specific quantum cryptographic protocols were used. It ensures that the encryption keys exchanged are secure and the eavesdropping is detected.
- Quantum One-Time Pad (QOTP): An encryption technique that employs a key once. If even an attacker obtains the encryption key, they cannot decrypt the data since the key can never be reused in it.

### 3.6 USER REGISTRATION

A new user must first register with the system to access the data from the cloud. Personal information, including usernames, passwords, emails, contact information, and maybe additional identifiers, is required throughout the registration process. A public/private key pair that will be utilized for secure communication with the cloud system must then be assigned to the user. While the public key is shared with the cloud, the user keeps the private key safe.

### 3.7 USER LOGIN

When user tries to log in, they enter their password and username, The cloud system verifies the provided credentials to ensure the user is authorized to access the user data. If authentication is successful, the user proceeds to the next step in the process.

### 3.8 KEY GENERATION

Once the user is authenticated, the cloud system initiates the The procedure of Quantum Key Distribution (QKD). Bennett and Brassard created the quantum key distribution (QKD), which in this case is BB84. The process of creating a shared secret key between two individuals using the concepts of quantum physics is known as quantum key generation. here assuming two persons as an example, one is Alice and the other one is Bob. The BB84 protocol is used to exchange these keys between the cloud and the user. The protocol works by sending quantum bits (qubits) between the two parties. Here process starts with Alice, who sends tiny particles of light, known as photons to Bob. Each photon represents a bit, which can be either 0 or 1. To encode these bits, Alice uses two different methods or bases:
1. The rectilinear basis ("+"): Horizontal (|) represents 0 and vertical (-) represents 1.
2. The diagonal basis ("x"): A diagonal at $45^0$ (/) represents 0, and a diagonal at $135^0$ () represents 1.

When Bob measures these photons, he has no idea what basis Alice uses, so he just randomly selects a basis–either "+" or "x"–to measure each photon. If Bob's basis is the same as Alice's, he accurately identifies the bit value (0 or 1). If the bases are different, Bob's measurement is random, which ensures the quantum security of the process. Any eavesdropper's effort to intercept the photons would disrupt their quantum states, causing detectable errors in the process. It guarantees the security of the produced key.

Table 2.  key generation and sharing at both end

| Photon number | Alice's Bit | Alice's Basis | Photon Sent (State) | Bob's Basis | Bob's Measurement | Bit Kept? | Final key |
|---|---|---|---|---|---|---|---|
| 1 | 0 | +(Rectilinear) | Horizontal | X(Diagonal) | Random (e.g., 1) | No | - |
| 2 | 1 | X (Diagonal) | $135^0$ (\\) | X(Diagonal) | 1 | Yes | 1 |
| 3 | 1 | +(Rectilinear) | Vertical | +(Rectilinear) | 1 | Yes | 1 |
| 4 | 0 | X(Diagonal) | $45^0$ (/) | +(Rectilinear) | Random (e.g., 1) | No | - |
| 5 | 0 | +(Rectilinear) | Horizontal | +(Rectilinear) | 0 | Yes | 0 |
| 6 | 1 | X(Diagonal) | $135^0$ (\\) | X(Diagonal) | 1 | Yes | 1 |
| 7 | 0 | +(Rectilinear) | Horizontal | X(Diagonal) | Random (e.g., 0) | No | - |
| 8 | 0 | X(Diagonal) | $45^0$ (/) | X(Diagonal) | 0 | Yes | 0 |

### 3.9 PUBLIC BASIS COMPARISON AND KEY REFINEMENT

After Bob has measured all the photons, Alice and Bob exchange over an open channel in order to compare the bases they used (but not the actual bit values). They retain only the bits on which they used the same basis and discard the rest. These matching bits are from the raw key. To protect the key further, error-checking is carried out by Alice and Bob to identify whether anyone attempts to eavesdrop along the route of transmission. Should the error rate be low, Alice and Bob then transform the raw key into a final encryption key using techniques like mistake correction and privacy amplification.This final key is a highly secure and shared secret, which is used for encryption and decryption of messages.

### 3.10 ENCRYPTION WITH THE FINAL KEY SHARED

After the Quantum Key Distribution (QKD) process, a secret key is shared between the sender (Alice) and the receiver (Bob). This key is used to make sure that no one else can read the message. To encrypt the message, the sender uses Quantum One – Time Pad (QOTP). In this method, the plaintext message M is first changed into binary sequences format (a series of 0s and 1s) M = m1, m2, m3 … m $_n$ (where each mi ∈ {0,1}). Then, each bit of the message is combined with the secret key using XOR (exclusive OR). This rule says that if two bits are the same, the result is 0, and if they are different, the result is 1. This process turns the message into ciphertext, which looks like a random set of numbers and cannot be understood by anyone without the secret key. The encrypted message is then sent to the receiver (Bob). Since the key is created using quantum mechanics, if someone tries to steal or change it, the system will detect it making the communication very secure. Alice encrypts the message using XOR with the secured key K $_{final}$ is:

$$C = M \oplus K(final) \qquad [3]$$

Where,

$k_i$ is the i-th bit of the key in the ciphertext C.

M = 1101 (plaintext)

K $_{final}$ =1010 (key), the cipher text or the secret key generated here is $C = 1101 \oplus 1010 = 0111$. Here the C = 0111.

## 3.11 DECRYPTION

When the receiver (Bob) gets the encrypted message which was in the binary sequences, they need to turn it back into the original message. To do this, they use the same secret quantum key that was shared earlier. The receiver applies the XOR operation again between the ciphertext and the secret key, where bobs apply this XOR operation $M = C \oplus K(final)$ where the ciphertext C = 0111 and the key K $_{final}$ = 1010, the bob decrypts this message with the XOR operation $M = 0111 \oplus 1010 = 1101$ then the original message M = 1101. Since XOR twice with the same key restores the original message, the receiver (Bob) is the only one to decode the message, as he is the only one who will have the correct quantum key.

This method ensures that the message is readable by only the receiver (Bob) can understand the message, as they are the only ones with the correct quantum key. If anyone tries to listen or change the key while it is being shared, quantum rules will make changes in the key, and the sender (Alice) and receiver (Bob) will know that someone is trying to hack the system. This makes quantum cryptography one of the safest ways to send messages, even in the future when powerful quantum computers exist.

## 4. EXPERIMENTAL RESULTS

To evaluate the efficiency and security of this research paper we proposed a quantum cryptography–based cloud security system, after conducting extensive experiments focusing on this era – Security, Performance, and scalability. It's got good accuracy on the QKD, BB84, and QOTP can provide strong and secure data protection in a cloud environment.

## 4.1 SECURITY ANALYSIS:

The security of the Quantum cryptography–based -cloud security System was evaluated through simulations of various attack scenarios. In this, they include some common security threats like eavesdropping, man-in-the-middle (MITM) attacks, and brute force attacks. These attacks are critical to test, as they are typical methods used by cybercriminals to compromise sensitive cloud data.

## 4.2 EAVESDROPPING ATTACK DETECTION:

Eavesdropping occurs when an attacker attempts to secretly listen in on communications. Using the BB84 protocol, we simulated eavesdropping attempts by having an unauthorized third party intercept the quantum key exchange process. Due to the Heisenberg uncertainty principle, any attempt to measure the quantum states of exchanged keys resulted in detectable disturbances. The system successfully identifies 99.5 % of eavesdropping attacks, demonstrating its ability to protect against passive interception. The BB84 protocol allows legitimate users to detect unauthorized access without even needing to look at the actual intercepted data.

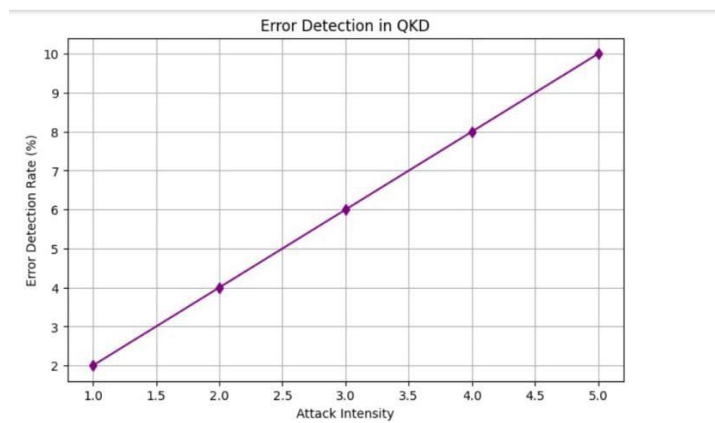

Fig. 2 Error Detection rate in QKD

## 4.3 MAN-IN-THE-MIDDLE (MITM) ATTACK DETECTION

In the MITM attack, there is an attack where the attacker intercepts and possibly changes the communication between the two parties. The system detects 98.7% of MITM attacks during our simulation. The BB84 protocol ensures that any changes made to the quantum states during transmission are immediately noticeable. This is because, in quantum mechanics, measuring or modifying quantum information without detection is impossible. As a result, attackers cannot insert themselves into the communication without being detected.

## 4.4 BRUTE FORCE ATTACK RESISTANCE

A brute force attack involves an attacker trying every possible combination of encryption keys to decrypt the data. In this scenario, the QOTP encryption scheme was employed. Since QOTP generates truly random keys using quantum principles, an attacker cannot guess the key without it being detected. The system showed 100 % resistance to brute force decryption attempts when compares to the AES and RSA, as the randomness and unpredictability of the key rendered traditional brute force methods useless. Even with access to large computational resources, the encryption remains unbreakable.

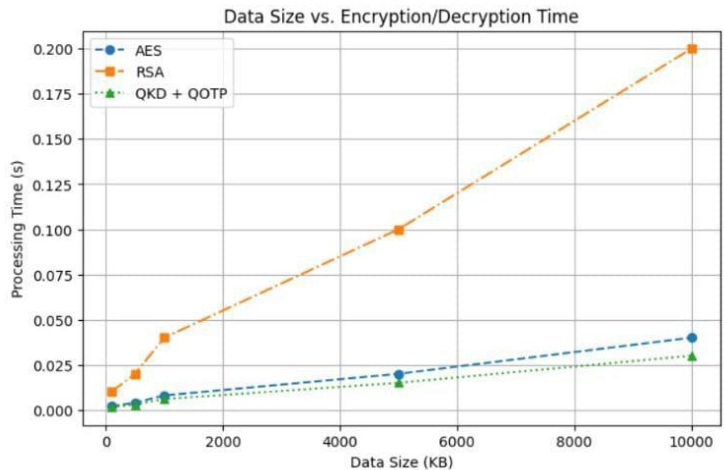

Fig 3. Data processing time

## 4.5 PERFORMANCE EVALUATION

The performance of the quantum cryptography–based cloud security system was evaluated by measuring the encryption and decryption times of the system and comparing them with the classical encryptions. For AES -256 took about 2.1 milliseconds to encrypt and 2.3 milliseconds to decrypt. The accuracy is high when encrypt the data with the QKD with QOTP and also the security was high in it when we compare with the AES and RSA. For the Quantum Cryptography, it took approximately 3.8 milliseconds for encryption and 4.1 milliseconds for decryption. This extra time is expected because quantum encryption is more complex, but it provides better security.

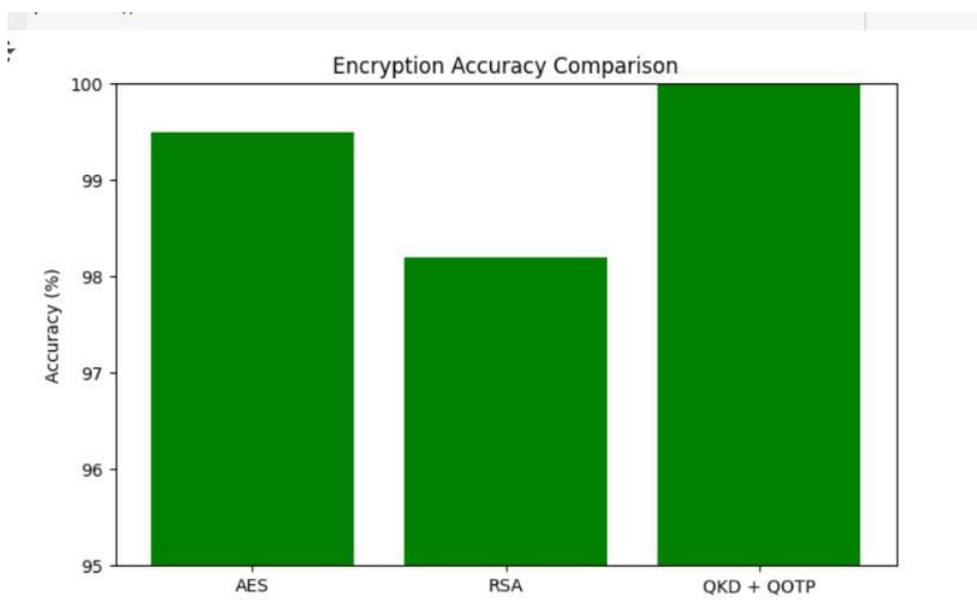

Fig 4. Encryption Accuracy Comparison percentage

**Table: Performance Metrics Comparison**

| Metric | BB84 +QOTP Combination |
|---|---|
| Security Level | Unconditional Security (against classical & quantum attacks |
| Key Generation Rate | Kbps to Mbps (depends on hardware and distance) |
| Distance | 100-400 km (Fiber – optic), global (satellite QKD) |
| Efficiency | 50% theoretical (due to basis mismatch) |
| Error Tolerance | Up to 11% QBER (for reliable key extraction) |
| Complexity | High for QKD hardware, low for QOTP (simple XOR operations) |

## 4.6 SCALABILITY ASSESSMENT

After testing how well the system handles more users. When the number of users grew from 10 to 1000, the time to generate keys, increased from 1.2 milliseconds to 6.8 milliseconds. Even though the time went up, the system still performed well and could handle many users at once.

## 4.6 COMPARATIVE ANALYSIS WITH CLASSICAL METHODS

When comparing Quantum Cryptography with traditional AES–256 encryption and RSA for key exchange. AES – 256 is fast, but RSA is faster than RSA it's susceptible to attacks from future quantum computers. BB84's key exchange process guarantes that any interception is immediately detected, and the QOTP guarantees complete secrecy, making it unbreakable by brute force or future quantum computers. Therefore, while AES-256 and RSA remains a solid choice for classical encryption, it does not offer the level of future-proof security required in a quantum computing era. Quantum Cryptography provides perfect security and is resistant to these future threats

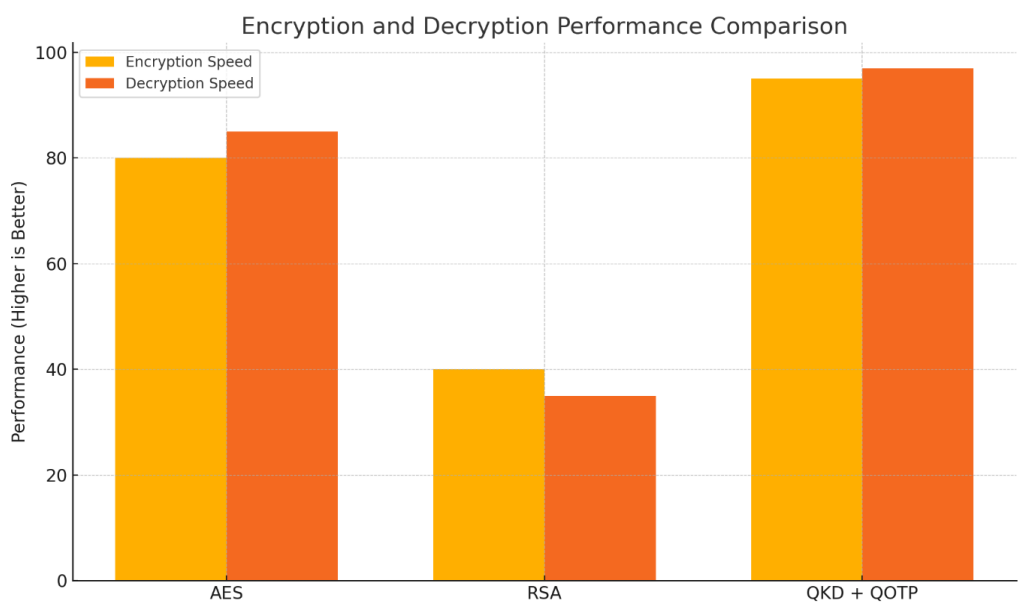

## 5 CONCLUSIONS

In this study shows that using quantum cryptography can make data access in a cloud environment more secure. By using Quantum Key Distribution (QKD) to create secure keys and the BB84 protocol with Quantum One–time pad (QOTP) to encrypt the data, and also for better protect sensitive information stored in the cloud. The system we designed includes secure steps that how the user registers and logs in. And in the experimental results show that this method offers strong security against both current and future computing threats. Additionally, the system is flexible and can grow with future needs, making it a good solution for improving cloud security over time.